\documentclass[3p,preprint]{elsarticle}
\usepackage{amssymb}
\usepackage{amsthm}
\usepackage{amsmath}
\usepackage[dvips]{epsfig}
\usepackage{graphicx}
\usepackage{color}
\usepackage{url}
\usepackage{bm,upgreek}
\usepackage{caption}
\usepackage{caption,setspace}
\usepackage{subcaption}
\usepackage{epstopdf}
\usepackage{multirow}
\usepackage{hhline}
\usepackage{slashbox}
\usepackage{cuted}

\begin{document}

\title{Localized modes in the Gross-Pitaevskii equation with a parabolic
	trapping potential and a nonlinear lattice pseudopotential}

\author[miet,ufa]{G.L. Alfimov\corref{gla}}
\ead{galfimov@yahoo.com}

\author[miet]{L.A. Gegel}
\author[ufa,viniti]{M.E. Lebedev}
\author[ta,itmo]{B.A. Malomed}
\author[itmo]{D.A. Zezyulin}
\ead{dzezyulin@corp.ifmo.ru}

\address[miet]{National Research University of Electronic Technology, Moscow, 124498, Russia}
\address[ufa]{Institute of Mathematics with Computer Center, Ufa Scientific Center, Russian Academy of Sciences, Chernyshevskii~str. 112, Ufa 450008, Russia}
\address[viniti]{All-Russian Institute for Scientific and Technical Information, Russian Academy of Sciences, 20 Usievich str, Moscow, 125190, Russia}
\address[ta]{Department of Physical Electronics, School of Electrical Engineering,
	Faculty of Engineering, Tel Aviv University, Tel Aviv 69978, Israel}
\address[itmo]{ITMO University, St.~Petersburg 197101, Russia}








\begin{abstract}
We study localized modes (LMs) of the one-dimensional
Gross-Pitaevskii/nonlinear Schr\"{o}dinger equation with a
harmonic-oscillator (parabolic) confining potential, 
and a periodically modulated coefficient in front of the cubic term
(nonlinear lattice pseudopotential). The equation applies to a cigar-shaped
Bose-Einstein condensate loaded in the combination of a magnetic trap and an
optical lattice which induces the periodic pseudopotential via the Feshbach
resonance. Families of stable LMs in the model feature specific properties
which result from the interplay between spatial scales introduced by the
parabolic trap and the period of the nonlinear pseudopotential. Asymptotic
results on the shapes and stability of LMs are  obtained for small-amplitude solutions and  in the limit  of a rapidly oscillating nonlinear
pseudopotential. We show that the presence of the lattice pseudopotential
may result in: (i) creation of new LM families which have no counterparts in
the case of the uniform nonlinearity; (ii) stabilization of some previously
unstable LM species; (iii) evolution of unstable LMs into a pulsating mode trapped
in one well of the lattice pseudopotential.
\end{abstract}


\maketitle
\section{Introduction}

Since the first production of the Bose-Einstein condensate (BEC) in
ultracold gases \cite{AEMWC95,DMADDKK95,BSTH95}, great progress has been
made in the experimental work with BEC, leading to the observation of
diverse species of matter waves, such as bright and dark solitons, gap
solitons, vortices, and other macroscopic quantum objects \cite{Romanian}.
In particular, apart from temporal or spatial variation of the trap that
confines the BEC, it is possible to control the scattering length (SL) of
interatomic interactions in BEC by means of the Feshbach-resonance (FR)
technique \cite{I98,TT04,KGJ06}. This technique allows one to change the
character of interatomic interactions, switching their sign from repulsive
to attractive one  and vice versa. Being compared to the size of the BEC
cloud, the characteristic scale of the spatial variation of SL may vary from
relatively large (e.g., if it is controlled by magnetic FR \cite{Hannaford}) to
moderate or small (for the optically-induced FR \cite{Japan,Cs}). The
interplay of the two characteristic scales, one being the trap's size and another one being  the scale of the SL variation, opens promising perspectives for the creation and
handling of novel stable matter-wave patterns \cite{HS}.

The commonly adopted mean-field model for BEC states is based on the
Gross-Pitaevskii equation (GPE) for the macroscopic wave function \cite%
{Pet,Pit}. This equation is, actually, a version of the classical nonlinear
Schr\"{o}dinger equation (NLSE) that takes into account a trapping
potential, $V(x)$, in the linear part of the equation, and a
spatially-dependent coefficient $P(x)$ in front of the cubic term (here,
one-dimensional settings are meant; in more general cases, the effective
one-dimensional nonlinearity, derived by the reduction of the
three-dimensional GPE, may assume algebraic forms, different from the simple
cubic term \cite{Salasnich,Delgado}). The latter $x$-dependent coefficient
defines a \textit{nonlinear pseudopotential }(with the name borrowed from
the theory of metals \cite{pseudo}), which is proportional to the local SL.
Usually, the experimentally relevant magnetic trap is modelled by the
harmonic-oscillator (HO), alias parabolic, potential. As concerns
pseudopotentials, various models have been used, including step \cite%
{ZAKP07,TAM16}, piecewise constant \cite{RKP08}, periodic \cite%
{SM05,WLK13,K17}, linear \cite{TSKF02}, Gaussian \cite{HY08}, and more
sophisticated well-shaped \cite{DM15} functions (for a comprehensive review
on the topic see \cite{KMT11}). In addition to pseudopotentials based on the
self-attractive nonlinearity, spatial modulation of the self-repulsive cubic
nonlinearity induces a new mechanism for the creation of self-trapped modes
in one- two-, and three-dimensional geometries \cite{Barcelona}.

In the present paper we consider the setting with the trapping potential in
the traditional HO form, and a spatially periodic nonlinear pseudopotential
(i.e., a \textit{lattice pseudopotential }\cite{KMT11}). This model
corresponds to atomic-gas BEC loaded in the combination of the HO-shaped
magnetic confinement and the optical lattice which periodically modifies the
local FR strength, cf. settings considered in Refs. \cite{WLK13,K17}, where
it was shown that the periodic modulation of the SL may result in
oscillatory instabilities of simplest dark solitons and stabilization of
more complex states \cite{K17}. It was also shown that the stability of
nonlinear modes depends on mutual position of the nonlinear pseudopotential
and the trapping potential \cite{WLK13}. However, those studies were chiefly
carried out for dark solitons, assuming strictly repulsive interatomic
interactions (in other words, solely positive SL), so that the nonlinear
pseudopotential $P(x)$ did not change its sign. To the best of our
knowledge, a comprehensive analysis of nonlinear localized modes (LMs) in
the HO potential, under the additional action of the periodic (generally,
sign-changing) variation of the SL, has not been reported previously.

The rest of the paper is organized as follows. The model is formulated in
Section~\ref{M_S}, which is followed by the analysis of small-amplitude LMs in
Section~\ref{SmAmpl}. In Section~\ref{NLM:general}, we proceed to the detailed numerical study of LMs of arbitrary amplitude, and Section~\ref{Concl} concludes the paper.

\section{The model and setup}

\label{M_S}

The effectively one-dimensional GPE\ for mean-field wave function $\Psi
\left( x,t\right) $, corresponding to the setting outlined above, is taken,
in a scaled form, as
\begin{equation}
i\Psi _{t}=-\Psi _{xx}+\frac{1}{2}\omega ^{2}x^{2}\Psi -P(x)\Psi |\Psi |^{2},
\label{1D_GP}
\end{equation}%
where $\omega ^{2}$ is the strength of the HO trapping potential, and the
nonlinear-lattice modulation function is periodic,%
\begin{equation}
P(x+2\pi /\Omega )=P(x).  \label{P(x)}
\end{equation}%
Intervals with positive (negative) values of $P(x)$ correspond to spatial
domains with attractive (repulsive) interactions between particles.
Equation~(\ref{1D_GP}) conserves two quantities, \textit{viz}., the energy,
\begin{equation}
E=\frac{1}{2}\int_{-\infty }^{+\infty }\left[ |\Psi_x|^{2}+\frac{1}{2}\omega ^{2}x^{2}|\Psi
|^{2}-\frac{P(x)}{2}|\Psi |^{4}\right] dx,  \label{N_E}
\end{equation}%
and the integral norm, which is proportional to the number of atoms in the
condensate:
\begin{equation}
N=\int_{-\infty }^{+\infty }|\Psi |^{2}dx.  \label{N_N}
\end{equation}

The model includes \textit{two scales}: the characteristic HO length $l_{%
	\mathrm{HO}}\sim 1/\sqrt{\omega }$, and period $T =2\pi /\Omega $ of
the nonlinear lattice, introduced by Eq. (\ref{P(x)}). In particular, the
limit case of the wide HO trap, $\Omega \gg 2\pi \sqrt{\omega }$, is a
physically relevant one. Additional rescaling
\begin{equation}
t\rightarrow \frac{\omega }{\sqrt{2}}t,\quad x\rightarrow \sqrt{\frac{\omega
	}{\sqrt{2}}}~x,\quad \Psi \rightarrow \sqrt{\frac{\sqrt{2}}{\omega }}~\Psi ,
\label{Rescal01}
\end{equation}%
allows one to fix $\omega \equiv 1$, converting Eq. (\ref{1D_GP}) into a
normalized form,
\begin{equation}
i\Psi _{t}=-\Psi _{xx}+x^{2}\Psi -P(x)\Psi |\Psi |^{2},  \label{1D_GP_1}
\end{equation}%
where $P(x)$ is periodic with spatial frequency $\tilde{\Omega}=\Omega /%
\sqrt{\omega }$ (in \textcolor{black}{what}  follows below, symbol $\tilde{\Omega}$ is replaced
by $\Omega $). To estimate physically relevant values of $\Omega $, we note
that, for the condensate of $^{173}$Yb used for the experimental realization
of the periodically-modulated FR in Ref. \cite{Japan}, the HO length
corresponding (for instance) to trapping frequency $\omega \sim 50$ Hz is $%
l_{\mathrm{HO}}\sim 5$ $\mathrm{\mu }$m, while the optical lattice was built
by laser beams with half-wavelength $\ell_{0}=0.278$ $\mathrm{\mu }$m,
the corresponding scaled spatial frequency being $\Omega \sim 100$. It can
be made smaller, taking larger $\omega $, and/or building the optical
lattice with larger $\ell =\ell_{0}/\cos \theta $, if the two laser
beams are launched under angle $\pi -2\theta $ \cite{OL}.

Equation~(\ref{1D_GP_1}) is our basic model. The objective of the analysis
is to construct and investigate stationary LM solutions to Eq. (\ref{1D_GP_1}%
) in the form of
\begin{equation}
\Psi (t,x)=e^{-i\mu t}u(x),
\end{equation}%
with real chemical potential $\mu $, and real \cite{AKS02} spatial wave
function $u(x)$ satisfying equation%
\begin{equation}
\frac{d^{2}u}{dx^{2}}+(\mu -x^{2})u+P(x)u^{3}=0  \label{eq1}
\end{equation}%
with localization boundary conditions,
\begin{equation}
\lim_{|x|\rightarrow \infty }u(x)=0.  \label{eq2}
\end{equation}

An important property of each LM is its stability with respect to small perturbations.  The linear stability of the LMs is addressed by taking
perturbed solutions in the form of \cite{JYang}
\begin{equation}
\textcolor{black}{\Psi (x,t)=\{u(x)+[v(x)+w(x)]e^{\lambda t}+[\bar{v}(x)-\bar{w}(x)]e^{\bar{\lambda}t}\}e^{-i\mu t}},
\end{equation}%
where $v(x)$ and $w(x)$ are infinitesimal perturbations, \textcolor{black}{and hereafter the  bar denotes   the complex conjugation.} Eigenvalues $%
\lambda $ with nonzero real parts give rise to the instability, while pure
imaginary ones correspond to linearly stable eigenmodes. Substituting this
expression in Eq.~(\ref{1D_GP_1}) and performing the linearization with
respect to $v$ and $w$, we arrive at the following linear eigenvalue problem
of the Bogoliubov -- de Gennes type,
\begin{equation}
\mathbf{L}\Theta =\lambda \Theta ,  \label{EigProb}
\end{equation}%
where
\begin{equation}
\mathbf{L}=i\left(
\begin{array}{cc}
\mathbf{0} & L_{-} \\
L_{+} & \mathbf{0}%
\end{array}%
\right) ,\quad \Theta =i\left(
\begin{array}{c}
v \\
w%
\end{array}%
\right)
\end{equation}%
\begin{align}
L_{-}& =d^{2}/dx^{2}+\mu -x^{2}+P(x)u^{2} \\[2mm]
L_{+}& =d^{2}/dx^{2}+\mu -x^{2}+3P(x)u^{2}.
\end{align}%
Note that, if $\lambda $ is an eigenvalue of Eq. (\ref{EigProb}), then $%
-\lambda $, $\bar{\lambda}$ and $-\bar{\lambda}$ are eigenvalues, too.

The eigenvalue problem based on Eq. (\ref{EigProb}) may be rewritten in the
following equivalent form:
\begin{equation}
L_{+}L_{-}w=\Lambda w,\quad \Lambda =-\lambda ^{2}.  \label{EigProb01}
\end{equation}%
In terms of Eq.~(\ref{EigProb01}), a mode $u(x)$ passes the linear stability test
if the spectrum of eigenvalues $\Lambda $ is all-real and positive.

\section{Small-amplitude localized modes: asymptotic analysis}

\label{SmAmpl}

\subsection{Branches of solutions}

\label{Gammas}

For small norm $N$, the nonlinear term in Eq. (\ref{eq1}) may be neglected,
which leads to the HO equation
\begin{equation}
\frac{d^{2}u}{dx^{2}}+(\mu -x^{2})u=0.  \label{LinGPE}
\end{equation}%
The latter  produces the commonly known set of eigenvalues and eigenfunctions:%
\begin{equation}
\tilde{\mu}_{n}=2n+1,\quad \tilde{u}_{n}(x)=\frac{1}{\sqrt{2^{n}n!\sqrt{\pi }%
}}H_{n}(x)e^{-\frac{1}{2}x^{2}},\quad n=0,1,\ldots ,  \label{eq:GH}
\end{equation}%
where $H_{n}(x)$ is $n$th Hermite polynomial. In particular,%
\begin{equation}
H_{0}(x)=1,\quad H_{1}(x)=2x,\quad H_{2}(x)=4x^{2}-2.  \label{012}
\end{equation}%
Eigenmodes $\tilde{u}_{n}(x)$ constitute an orthonormal basis in $L_{2}(%
\mathbf{R})$,
\begin{equation}
\langle \tilde{u}_{n},\tilde{u}_{m}\rangle \equiv \int_{-\infty }^{\infty }%
\tilde{u}_{n}(x)\tilde{u}_{m}(x)~dx=\delta _{m,n}.
\end{equation}

When the nonlinearity is switched on, each linear eigenstate $(\tilde{\mu}%
_{n},\tilde{u}_{n}(x))$ bifurcates into a one-parameter set $\Gamma
_{n}=(\mu _{n},u_{n}(x))$ of small-amplitude LMs. These modes are produced
by Eq.~(\ref{eq1}) and become essentially nonlinear with the increase of $N$
(in other words, with the increase of the distance between $\mu _{n}$ and $%
\tilde{\mu}_{n}$). Following the terminology adopted in Refs. \cite{AMP,AP},
we refer to these solutions as to \textit{nonlinear modes with linear
	counterparts}. LMs from family $\Gamma _{n}$ feature the same parity as the
corresponding linear eigenfunction $\tilde{u}_{n}(x)$: the LMs with even $n$
are even functions of $x$, and those with odd $n$ are odd. LMs belonging to branch $\Gamma _{0}$,
which originates from the HO ground state, are nodeless, resembling the
well-known bright solitons of the NLSE with the attractive nonlinearity.
The LMs belonging to branch $\Gamma _{1}$, which originates from the first
HO excited state have exactly one node, somewhat resembling dark solitons of
the repulsive NLSE (these solutions are also similar to the so-called
localized dark solitons of the NLSE with the strength of the local
self-repulsion slowly growing from $x=0$ towards $|x|\rightarrow \infty $
\cite{Zeng}).
They have been studied in many papers, see, e.g. \cite{WLK13,K17,C10,PFK05}.

If the amplitude of $u_{n}(x)$ is small, the LMs from branch $\Gamma _{n}$
are approximated by expansions \cite{K17,KKRF05,ZAKP08,ZK16},
\begin{equation}
u_{n}(x)=\varepsilon \tilde{u}_{n}(x)+o(\varepsilon ),\quad \mu _{n}=\tilde{%
	\mu}_{n}-\varepsilon ^{2}\Delta _{n}+o(\varepsilon ^{2}),  \label{AsEx}
\end{equation}%
where $\varepsilon \ll 1$ is a small parameter, and%
\begin{equation}
\Delta _{n}=\int_{-\infty }^{+\infty }P(x)\tilde{u}_{n}^{4}(x)dx=\frac{1}{%
	2^{2n}(n!)^{2}\pi }\int_{-\infty }^{+\infty }P(x)H_{n}^{4}(x)e^{-2x^{2}}~dx.
\label{omega2}
\end{equation}

\subsection{The stability of small-amplitude localized modes}

\label{Stab_Gamma}

To address the stability of the small-amplitude nonlinear LMs belonging to
branches $\Gamma _{n}$, $n=0,1,\ldots $, we note that, with the help of
expansion (\ref{AsEx})--(\ref{omega2}), operator $L_{+}L_{-}$ in (\ref%
{EigProb01}) may be considered as a perturbation of operator $\mathcal{L}%
_{n}^{2}$, where
\begin{equation}
\mathcal{L}_{n}=\frac{d^{2}\,}{dx^{2}}+2n+1-x^{2}.
\end{equation}%
Specifically,
\begin{eqnarray}
&&L_{+}=\mathcal{L}_{n}+\varepsilon ^{2}(3P(x)\tilde{u}_{n}^{2}(x)-\Delta
_{n})+o(\varepsilon ^{2}), \\[2mm]
&&L_{-}=\mathcal{L}_{n}+\varepsilon ^{2}(P(x)\tilde{u}_{n}^{2}(x)-\Delta
_{n})+o(\varepsilon ^{2}), \\[2mm]
&&L_{+}L_{-}=\mathcal{L}_{n}^{2}+\varepsilon ^{2}M_{n}+o(\varepsilon ^{2}),
\label{eq:LL}
\end{eqnarray}%
where
\begin{equation}
M_{n}=(3P(x)\tilde{u}_{n}^{2}(x)-\Delta _{n})\mathcal{L}_{n}+\mathcal{L}%
_{n}(P(x)\tilde{u}_{n}^{2}(x)-\Delta _{n}).
\end{equation}%
Operator $\mathcal{L}_{n}$ is self-adjoint in $L^{2}$, and its spectrum
consists of eigenvalues $\varkappa _{k}=2(n-k)$ with corresponding
eigenfunctions $\tilde{u}_{k}(x)$, $k=0,1,\ldots $, see (\ref{eq:GH}). The spectrum is
equidistant, all the eigenvalues being simple. There are infinitely
many negative eigenvalues, $n$ positive eigenvalues, and one zero eigenvalue. The
eigenvalues of the operator $\mathcal{L}_{n}^{2}$ are squared eigenvalues of
$\mathcal{L}_{n}$, $\tilde{\Lambda}_{k}=\varkappa _{k}^{2}=4(k-n)^{2}$,
corresponding to the same eigenfunctions $\tilde{u}_{k}(x)$, $k=0,1,\ldots $. This means that the spectrum of $\mathcal{L}_{n}^{2}$ includes $n$ \textit{%
	double positive eigenvalues} $\tilde{\Lambda}_{k}=4(n-k)^{2}$, $k=0,1,\ldots
(n-1)$, one \textit{simple zero eigenvalue} and \textit{infinitely many
	simple positive eigenvalues}. Each of the double eigenvalues has an invariant
subspace spanned by two functions, $\tilde{u}_{k}(x)$ and $\tilde{u}%
_{2n-k}(x)$. If $n=0$, then all eigenvalues of $\mathcal{L}_{n}^{2}$ are simple.

Generically, small perturbation of $\mathcal{L}_{n}^{2}$ results in \textit{splitting} of the double eigenvalues. Each of them can split   into (i)
two real eigenvalues of the perturbed operator or (ii) two complex-conjugate
eigenvalues. If the case~(i) takes place for each double eigenvalue, then small-amplitude LM bifurcating from the $n$th linear eigenstate are marginally stable, at least in some vicinity of the bifurcation. However, if at least for one double eigenvalue the case~(ii) takes place, then the bifurcating small-amplitude LMs are unstable in some vicinity of the bifurcation.
%

To address the splitting of double eigenvalues when passing from operator $%
\mathcal{L}_{n}^{2}$ to the perturbed one, $L_{+}L_{-}$ in Eq.~(\ref{eq:LL}%
), we construct an asymptotic expansion for perturbed eigenvalues following  Ref.~\cite{ZAKP08}. Under the action of
the perturbation, each double eigenvalue $\tilde{\Lambda}$ splits into two
simple ones:
\begin{equation}
\Lambda _{1}=\tilde{\Lambda}+\varepsilon ^{2}\gamma _{1}+o(\varepsilon
^{2}),\quad \Lambda _{2}=\tilde{\Lambda}+\varepsilon ^{2}\gamma
_{2}+o(\varepsilon ^{2}),
\end{equation}%
where the coefficients $\gamma _{1,2}$ are the eigenvalues of the $2\times 2$ matrix
\begin{equation}
\tilde{M}_{n}=\left(
\begin{array}{cc}
\langle M_{n}\tilde{u}_{k},\tilde{u}_{k}\rangle & \langle M_{n}\tilde{u}_{k},%
\tilde{u}_{2n-k}\rangle \\[2mm]
\langle M_{n}\tilde{u}_{2n-k},\tilde{u}_{k}\rangle & \langle M_{n}\tilde{u}%
_{2n-k},\tilde{u}_{2n-k}\rangle%
\end{array}%
\right).  \label{Lambda_12_as}
\end{equation}%
Therefore, if the eigenvalues of $\tilde{M}_{n}$ are real for each $k=0,1,\ldots, n-1$, then the spectrum of $%
L_{+}L_{-}$ remains real and the nonlinear LM $u_{n}(x)$ is stable, at least
for sufficiently small $\varepsilon $ (i.e., for sufficiently small number
of particles in the condensate). Otherwise, if eigenvalues of $\tilde{M}$
are complex  for some  $k=0,1,\ldots, n-1$, then the LM solution $u_{n}(x)$ is unstable in a vicinity of the bifurcation which gives rise to the complex eigenvalue pair. Note that, as
no double eigenvalues exist in the case of $n=0$, the small-amplitude LMs
belonging to the ground-state branch $\Gamma _{0}$ are stable for any $P(x)$.

Using explicit expression for the eigenfunction $\tilde{u}_{n}$ from (\ref{eq:GH}%
), one can compute the entries of the matrix $\tilde{M}_{n}$:
\begin{align}
\langle M_{n}\tilde{u}_{k}& ,\tilde{u}_{k}\rangle =\frac{8(n-k)}{\pi
	2^{(n+k)}n!k!}\int_{-\infty }^{+\infty
}P(x)H_{n}^{2}(x)H_{k}^{2}(x)e^{-2x^{2}}~dx   \notag \\[2mm]
& -\frac{4(n-k)}{\pi 2^{2n}(n!)^{2}}\int_{-\infty }^{+\infty
}P(x)H_{n}^{4}(x)e^{-2x^{2}}~dx,  \label{11} \\[2mm]
\langle M_{n}\tilde{u}_{k}& ,\tilde{u}_{2n-k}\rangle =-\langle M_{n}\tilde{u}%
_{2n-k},\tilde{u}_{k}\rangle =  \notag \\[2mm]
& =\frac{4(n-k)}{\pi 2^{2n}n!\sqrt{k!(2n-k)!}}\int_{-\infty }^{+\infty
}P(x)H_{n}^{2}(x)H_{2n-k}(x)H_{k}(x)e^{-2x^{2}}~dx,  \label{12} \\[2mm]
\langle M_{n}\tilde{u}_{2n-k}& ,\tilde{u}_{2n-k}\rangle =-\frac{8(n-k)}{\pi
	2^{(3n-k)}n!(2n-k)!}\int_{-\infty }^{+\infty
}P(x)H_{n}^{2}(x)H_{2n-k}^{2}(x)e^{-2x^{2}}~dx  \notag \\[2mm]
& +\frac{4(n-k)}{\pi 2^{2n}(n!)^{2}}\int_{-\infty }^{\infty
}P(x)H_{n}^{4}(x)e^{-2x^{2}}~dx.  \label{22}
\end{align}

Formulas (\ref{11})-(\ref{22}) with $P(x)=\pm 1$ were used in Ref. \cite%
{ZAKP08} to explore the stability of small-amplitude nonlinear modes 
in the model with constant scattering length.

\section{Branches of nonlinear localized modes: a numerical study}

\label{NLM:general}

We present numerical results for the practically important case when the
nonlinearity-modulation function, $P(x)$ in Eq.~(\ref{1D_GP_1}), is taken as
a sum of its constant (dc) and harmonic (ac) parts:
\begin{equation}
P(x)=P_{0}+P_{1}\cos \left( \Omega x\right) .  \label{TwoH_App}
\end{equation}%
In what follows we conclude (quite naturally) that the relation between the
magnitudes of $|P_{0}|$ and $|P_{1}|$ is important, hence it is necessary to
consider two cases separately: (a) $|P_{0}|\gtrsim |P_{1}|$ (the dc component is not negligible,
the {\it dc-ac case}); (b) $|P_{0}|\ll |P_{1}|$ (the dc component is negligible,
the {\it ac case}).

In the \textit{dc-ac case} (a) one can scale out the absolute value
of the dc component, by replacing
\begin{equation}
{\Psi \rightarrow \Psi /\sqrt{|P_{0}|},~} \quad P_{1}/|P_{0}|\rightarrow P_{1},
\label{Rescal02}
\end{equation}%
and thus casting Eq.~(\ref{1D_GP_1}) in the form of
\begin{equation}
i\Psi _{t}=-\Psi _{xx}+x^{2}\Psi {-}\left[ \sigma _{0}+P_{1}\cos \left(
\Omega x\right) \right] \Psi |\Psi |^{2},\quad {\sigma _{0}\equiv
	P_{0}/|P_{0}|=\mathrm{sign}~P_{0}.}  \label{1D_GP_11}
\end{equation}

In the \textit{ac case} (b), we drop $P_{0}$, and rescale Eq. (\ref{1D_GP_1}) by replacing
\begin{equation}
{\Psi \rightarrow \Psi /\sqrt{|P_{1}|}},
\end{equation}%
which leads to the equation
\begin{equation}
i\Psi _{t}=-\Psi _{xx}+x^{2}\Psi {-}\sigma _{1}\cos \left( \Omega x\right)
\Psi |\Psi |^{2},\quad \sigma _{1}=P_{1}/|P_{1}|=\mathrm{sign}~P_{1}.
\label{1D_GP_12}
\end{equation}%
%
%

\subsection{Nonlinear pseudopotential with nonzero mean (dc case)}

\label{NonZM}

LMs provided by Eq.~(\ref{1D_GP_11}) satisfy the equation
\begin{equation}
\frac{d^{2}u}{dx^{2}}+(\mu -x^{2})u+\left[ \sigma _{0}+P_{1}\cos \left(
\Omega x\right) \right] u^{3}=0,\quad \sigma _{0}=\mathrm{sign}~P_{0}.
\label{eq1_1}
\end{equation}

\subsubsection{The constant pseudopotential: an overview}

\label{Review_const}

\begin{figure}[tbp]
	\begin{center}
	\includegraphics[width=0.8\textwidth]{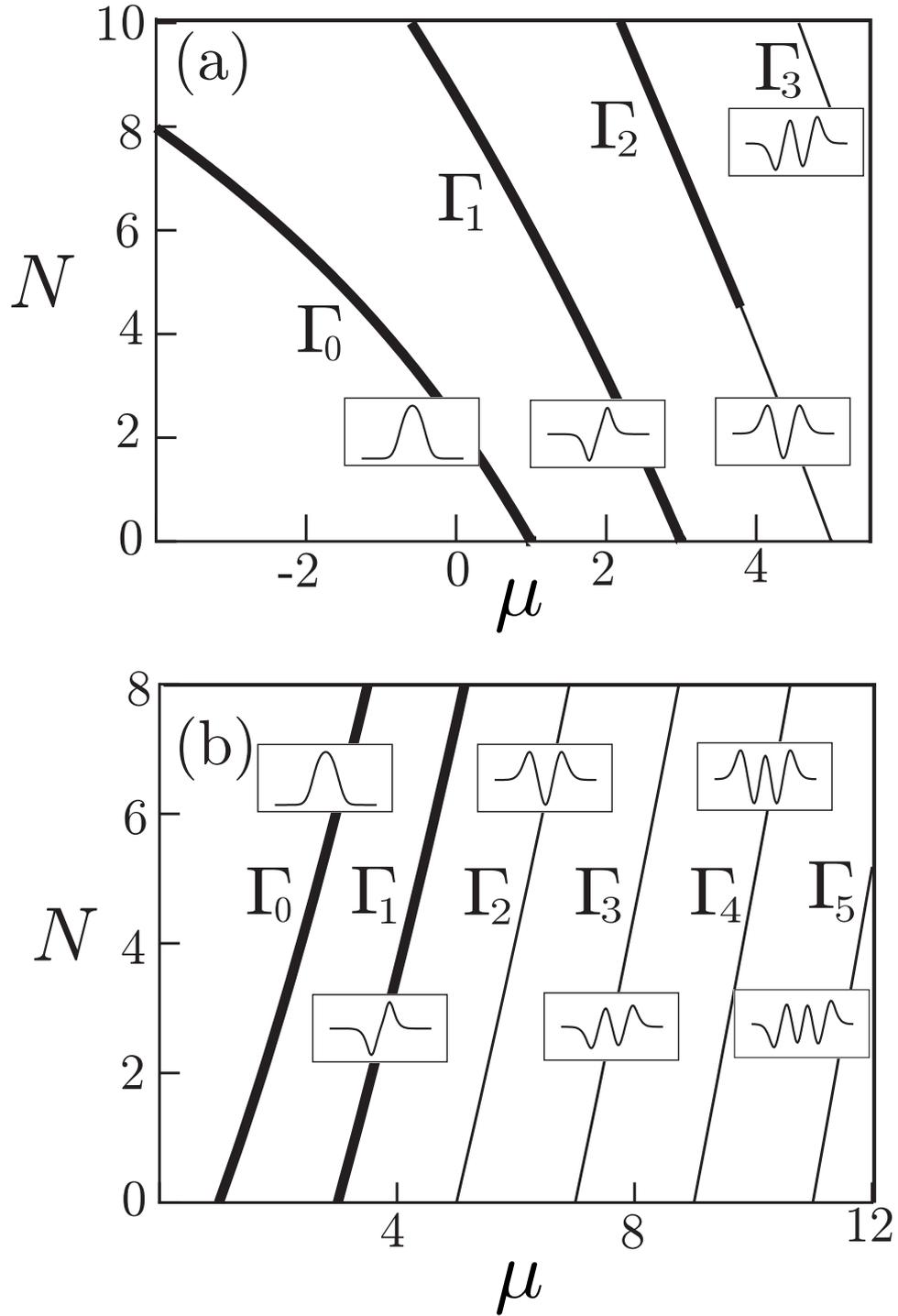}
	\caption{$N(\protect\mu )$ curves for the NHO model, based on Eq.~(\protect
		\ref{eq1_1_0}), with (a) $\protect\sigma _{0}=1$ and (b) $\protect\sigma %
		_{0}=-1$, which correspond, respectively, to the self-attractive and
		repulsive signs of the nonlinearity (as per Refs. \protect\cite%
		{ZAKP07,ZAKP08}). Bold (thin) segments correspond to stable (unstable)
		modes. Insets show schematic profiles $u_{n}(x)$ of modes from each family.}
	\label{Fig.P_1=1}
	\end{center}
\end{figure}

Before presenting new results produced by the current work, it is relevant
to briefly overview the previous results pertaining to the well-studied case of
the GPE with constant (negative or positive) scattering length \cite%
{C10,ZAKP08,E95,K99, G99,  KAT01, C01, YYB02, KK03,AZ07}. In this case, $%
P_{1}\equiv 0$, and Eq.~(\ref{eq1_1}) becomes
\begin{equation}
\frac{d^{2}u}{dx^{2}}+(\mu -x^{2})u+\sigma _{0}u^{3}=0.  \label{eq1_1_0}
\end{equation}%
We call Eq.~(\ref{eq1_1_0}) the \textit{nonlinear harmonic-oscillator equation}%
. The case of $\sigma _{0}=1$ ($\sigma _{0}=-1$) corresponds to the
attractive (repulsive) interparticle interactions. It is convenient to
illustrate the branches of LMs in the nonlinear HO model by means of the
respective $N(\mu )$ curves, which are presented in Fig. \ref{Fig.P_1=1}, as per Ref. \cite{ZAKP07,ZAKP08}, for both cases of $\sigma _{0}=\pm 1$.\
The branches $\Gamma _{n}$, $n=0,1,2,\ldots $, correspond to the LMs with the
linear counterparts, bifurcating from them at the points $\tilde{\mu}_{n}=2n+1$,
$N=0$, all the branches being represented by monotonous functions $N(\mu )$
(at least, for moderate values of $N$, $\mu $ and $n$) . Presumably, there
exist no nonlinear HO modes without linear counterparts \cite{AZ07}.

Analysis of the eigenvalues of the matrix $\tilde{M}_{n}$ with $P(x)\equiv 1$ or
$P(x)\equiv -1$ \cite{ZAKP08} shows that LMs corresponding to $\Gamma _{0}$
and $\Gamma _{1}$ are stable in the small-amplitude limit for both signs of
the nonlinearity, $\sigma _{0}=1$ and $\sigma _{0}=-1$ \cite{KKRF05,AZ07}.
Numerical results indicate that these modes remain stable for moderate and
large amplitudes as well. The small-amplitude LMs belonging to the branch $\Gamma _{2}$ are unstable. For $\sigma _{0}=1$ (self-attraction), the
instability of the branch $\Gamma _{2}$ persists for $\mu ^{\ast }<\mu <5$ where $\mu ^{\ast }\approx 3.83$. At $\mu <\mu ^{\ast }$ these modes was reported
to be stable \cite{AZ07}. For $\sigma _{0}=-1$ (self-repulsion), the branch $\Gamma _{2}$ are unstable in the whole interval of $\mu $ covered in Fig.~%
\ref{Fig.P_1=1} (however, it may become stable at still larger values of $N,$
not shown in Fig.~\ref{Fig.P_1=1} \cite{C10}). The small-amplitude modes for the branches $%
\Gamma _{n}$ with $n=3,4,\ldots $ are also unstable in both cases of $\sigma
_{0}=\pm 1$.

\subsubsection{The dc-ac case: an oscillating pseudopotential}

\label{P1_ne_0}

\begin{figure}[tbp]
	\begin{center}
	\includegraphics[width=0.7\textwidth]{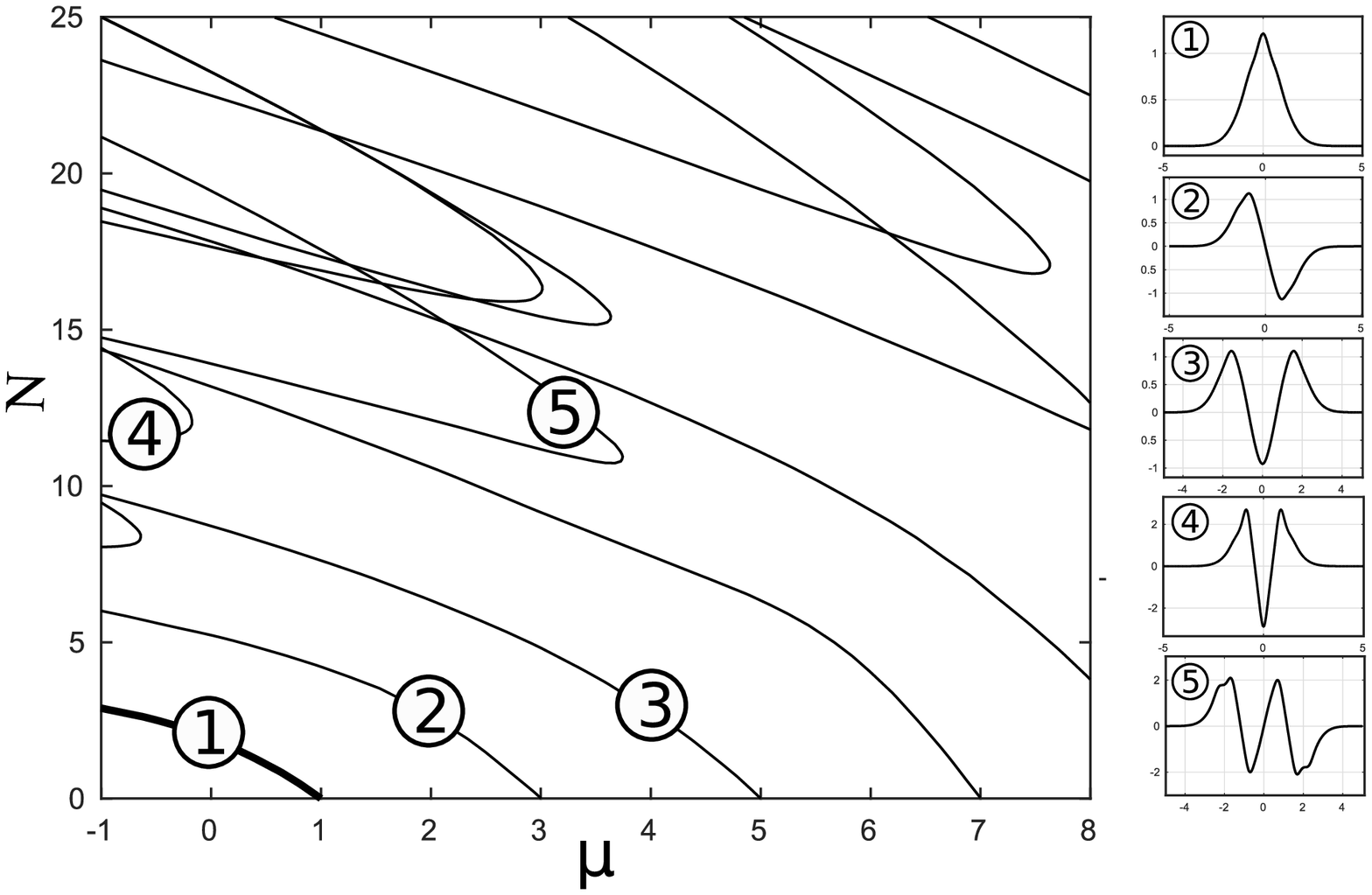}
	\caption{The dc-ac case. The main panel: $N(\protect\mu )$ dependences, as obtained from
		Eq.~(\protect\ref{eq1_1}) with $\protect\sigma _{0}=1$, $P_{1}=2$, and $%
		\Omega =8$. Thin and bold lines show unstable and stable LM families, the
		sole stable one being the lowest branch $\Gamma _{0}$. Insets (1)-(3) are
		representative profiles $u(x)$ of nonlinear modes originating their linear
		counterparts, for branches $\Gamma _{0}$, $\Gamma _{1}$, $\Gamma _{2}$
		(labeled by the same numbers $1,2,3$), while insets (4) and (5) show
		profiles of modes belonging to some branches without linear counterparts,
		which are also labeled $4$ and $5$. }
	\label{FigP_1ne_0_1}
\end{center}
\end{figure}

Let us now   consider the combination of the HO
trapping potential with the pseudopotential which contains the periodic
component, i.e., $P_{1}\neq 0$ in Eq. (\ref{eq1_1}). In this case, a general
picture of LM branches can be obtained by means of a numerical shooting
algorithm, which is presented in detail in Ref. \cite{AZ07}.
A representative example of the respective $N(\mu )$ curves
with $\sigma _{0}=1$ (average self-attraction), $P_{1}=2$ (which implies
that the sign of the local nonlinearity periodically flips) and $\Omega =8$
is displayed in Fig.~\ref{FigP_1ne_0_1}, where one immediately observes
that, apart from the branches $\Gamma _{n}$ originating from their linear
counterparts, numerous branches of LMs \emph{without linear counterparts}
exist too. Thus, the presence of the ac component in the pseudopotential
essentially enriches the diversity of available solutions. However, no
branch without a linear counterpart is stable (in fact, the only stable
solutions in Fig.~\ref{FigP_1ne_0_1} corresponds to the ground-state branch,
$\Gamma _{0}$).

In the limit of the rapidly oscillation ac component, $\Omega \rightarrow
\infty $, LMs 
may be approximated by the nonlinear HO modes. The asymptotic formula can be
obtained by means of averaging with respect to the fast oscillations (see,
e.g., Ref. \cite{Fatkhulla}):
\begin{equation}
u(x)=u^{(0)}(x)+\frac{1}{\Omega ^{2}}\left[ u^{(1)}(x)+P_{1}(u^{(0)}(x))^{3}%
\cos \left( \Omega x\right) \right] +o\left( \frac{1}{\Omega ^{2}}\right),\quad \Omega \rightarrow \infty.  \label{Expand_NZM}
\end{equation}%
Here $u^{(0)}(x)$ is a solution of nonlinear HO model corresponding to Eq. (%
\ref{eq1_1_0}), and $u^{(1)}(x)$ is a localized solution of the linear equation
\begin{equation}
\frac{d^{2}u}{dx^{2}}+\left\{ \mu -x^{2}+3\sigma _{0}\left[ u^{(0)}(x)\right]
^{2}\right\} u=-\frac{3}{2}P_{1}^{2}\left[ u^{(0)}(x)\right] ^{5}.
\end{equation}%
{We stress that asymptotic relation (\ref{Expand_NZM}) is valid for
	nonlinear modes of arbitrary amplitudes (i.e., not only in the
	small-amplitude limit), provided that $\Omega ^{2}$ is large enough.}

\begin{figure}[tbp]
	\begin{center}
	\includegraphics[width=0.8\textwidth]{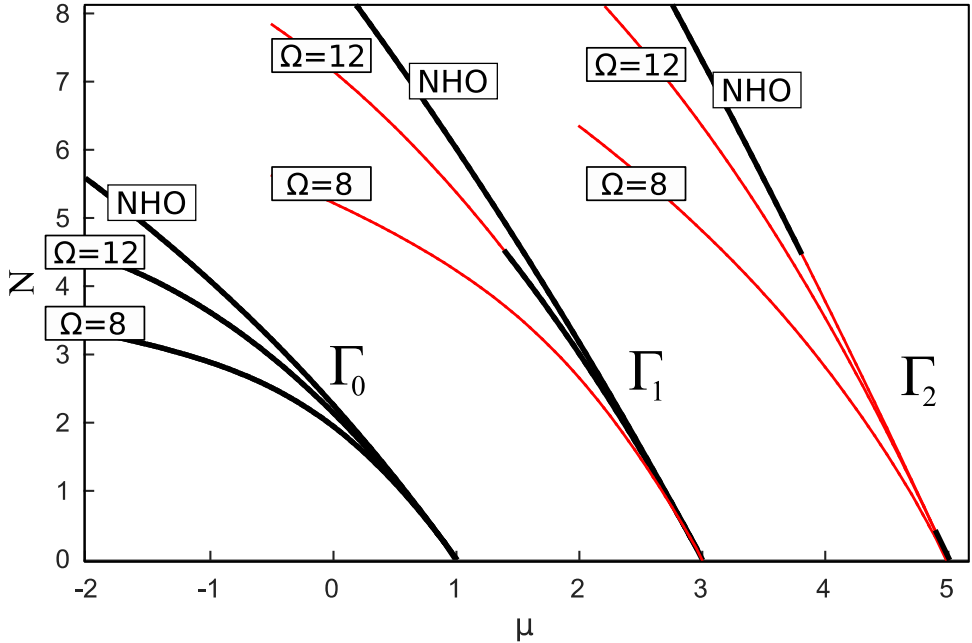}
	\caption{The dc-ac case. $N(\protect\mu )$ curves for Eq.~(\protect\ref{eq1_1}) with $%
		\protect\sigma _{0}=1$, $P_{1}=2$, and $\Omega =8$ or $\Omega =12$. Only
		branches $\Gamma _{0,1,2}$ are shown. For comparison, the corresponding
		dependencies for NHO equation (\protect\ref{eq1_1_0}) with $\protect\sigma %
		_{0}=1$ [identical to those in Fig.~\protect\ref{Fig.P_1=1}(a)] are shown
		too. Bold black lines (and thin red lines) correspond to stable (unstable)
		nonlinear modes.}
	\label{FigP_1ne_0}
\end{center}
\end{figure}

Figure~\ref{FigP_1ne_0} shows the branches $\Gamma _{0}$, $\Gamma _{1}$ and $%
\Gamma _{2}$ for 
two different spatial frequencies of the periodic pseudopotential, $\Omega =8
$ and $\Omega =12$. According to the asymptotic prediction (\ref{Expand_NZM}%
), they approach the corresponding branches of the nonlinear HO equation as $%
\Omega $ grows. Additionally, Eq. (\ref{Expand_NZM}) implies that ({in)}%
stability of a LM under the action of the rapidly oscillating
pseudopotential is determined by the ({in)}stability of its counterpart in
the nonlinear HO model. {Indeed, the presence of eigenvalues with nonzero
	real parts in the perturbative spectrum of an nonlinear HO state implies the
	existence of such eigenvalues in the spectrum of LM if $\Omega $ is large
	enough. We also conjecture that the stability of a nonlinear HO mode implies
	the stability of its LM counterpart for sufficiently large $\Omega $.}
For example, the segment of the branch $\Gamma _{1}$ shown in Fig.~\ref%
{FigP_1ne_0_1} is completely unstable for $\Omega =8$, but its stability
restores for $\Omega =12$, which agrees with the nonlinear HO limit, where
this branch is entirely stable. For the branch $\Gamma _{2}$ the situation is
more complex. Small-amplitude modes belonging to $\Gamma _{2}$ are unstable
for $\Omega =8$, but become stable for $\Omega =12$. However, Eq. (\ref%
{Expand_NZM}) implies that the further increase of $\Omega $ will
necessarily lead to the destabilization of these modes, because in the
nonlinear HO limit the small-amplitude modes belonging to $\Gamma _{2}$ are
unstable. The possibility to manage the stability of small-amplitude
nonlinear LMs by tuning the frequency of the ac component of the
pseudopotential has been recently reported in Ref. \cite{K17}.

\begin{figure}[tbp]
	\begin{center}
	\includegraphics[width=0.7\textwidth]{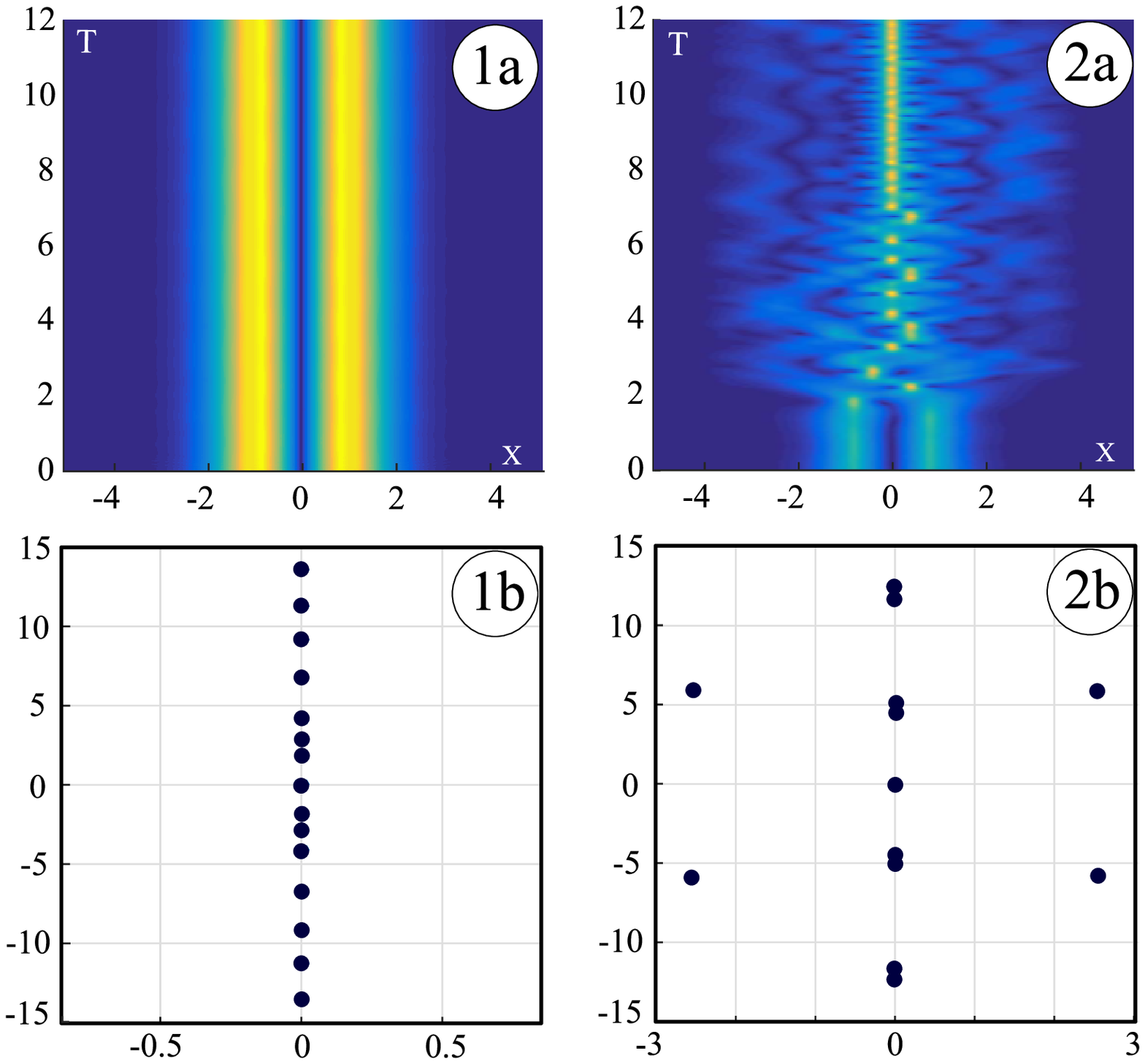}
	\caption{The dc-ac case, $P(x)=1+2\cos 16 x$.
		(1a): The
		evolution of a stable LM with $\protect\mu =0$, belonging to branch $\Gamma
		_{1}$. (1b): The spectrum of the eigenvalue problem (\protect\ref{EigProb})
		associated with this LM. (2a): The evolution of an unstable LM with $\protect%
		\mu =-3$, belonging to branch $\Gamma _{1}$. (2b) The corresponding spectrum
		of the eigenvalue problem (\protect\ref{EigProb}). }
	\label{Fig_Comp_NZM}
	\end{center}
\end{figure}

To check the predictions of the linear-stability analysis, we have performed
simulations of the evolution of LMs in the framework of the time-dependent
GPE~(\ref{1D_GP_11}), using an implicit finite-difference scheme \cite{TP} with
adsorbing boundary conditions. In the simulations, solutions which are
predicted to be linearly stable keep their shape indefinitely long [see Fig.~
\ref{Fig_Comp_NZM}~(1a,b)], whereas the nonlinear modes which are predicted
to be unstable typically transform into a pulsating object localized over
one period of the lattice pseudopotential, see Figs.~\ref{Fig_Comp_NZM}~(2a,b).

\subsection{Nonlinear pseudopotential with zero mean (the ac case)}

\label{ZeroMean}

\begin{figure}[tbp]
	\begin{center}
	\includegraphics[width=0.9\textwidth]{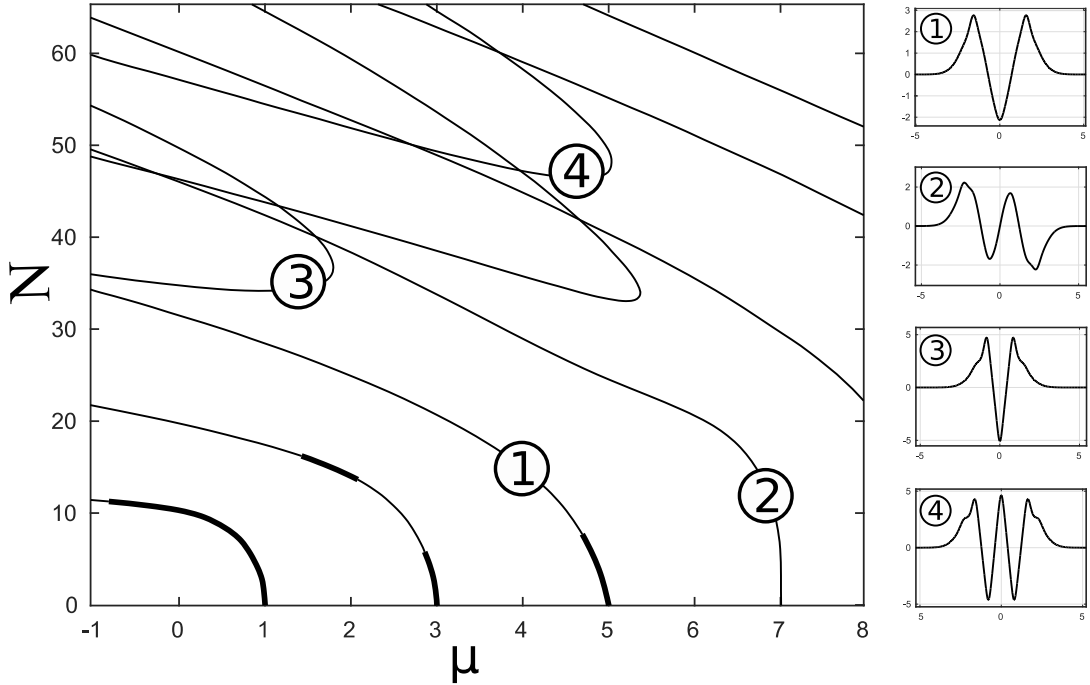}
	\caption{The ac case. The main panel: $N(\protect\mu )$ curves for $P(x)=\cos (8x)$, thin
		and bold lines showing unstable and stable LM families, respectively. (1)
		and (2) label branches $\Gamma _{2}$, $\Gamma _{3}$, while \textcolor{black}{(3) and (4)}
		pertain to ones without linear counterparts. Profiles of the LMs for
		branches (1)-(4) are depicted in separate panels.}
	\label{A0Om8_l}
	\end{center}
\end{figure}

According to the results of the previous subsection, the effect of a rapidly
oscillating ac component of the pseudopotential, in the presence of the dc
component, may be approximated using the standard nonlinear HO model with
the uniform nonlinearity. However, the situation becomes essentially
different in the absence of the dc component. We address the latter
situation using Eq.~(\ref{1D_GP_12}). The corresponding nonlinear modes can
be found from the equation
\begin{equation}
\frac{d^{2}u}{dx^{2}}+(\mu -x^{2})u+\sigma _{1}\cos (\Omega x)u^{3}=0.
\label{eq1_2}
\end{equation}%
The so obtained $N(\mu )$ curves are shown in the main panel of Fig.~\ref%
{A0Om8_l}. Here we again observe the branches $\Gamma _{n}$ bifurcating from the
linear limit [Fig. \ref{A0Om8_l}, insets (1) and (2)] and various families
without linear counterparts [Fig.~\ref{A0Om8_l}, \textcolor{black}{insets (3-4)}]. The branches $%
\Gamma _{n}$ feature stable and unstable segments, all LMs without linear
counterpart being unstable, as above.

\subsubsection{Shapes of solutions with linear counterparts}

\label{Gamma_shapes}

In the limit $N\ll 1$, weakly nonlinear LMs belonging to branches $\Gamma
_{n}$, $n=0,1,\ldots $, are produced by Eqs. (\ref{AsEx}) where $\Delta _{n}$
is given by Eq. (\ref{omega2}), with $P(x)=\sigma _{1}\cos \left( \Omega
x\right) $. The numerically found values of $\Delta _{n}$, $n=0,1,2$, for $%
\sigma _{1}=1$ and several different values of $\Omega $ are presented in
Table~\ref{T2}.

\begin{table}[tbp]
	\begin{center}
	\begin{tabular}{ccccc}
		\hline
		$\Omega$ & 0 & 2 & 8 & 18 \\[2mm] \hline
		$n=0$ & $\quad 3.9894\cdot 10^{-1}$ \quad & \quad $2.4197\cdot 10^{-1}$ \quad
		& \quad $1.3383\cdot 10^{-4}$ \quad & \quad $1\cdot 10^{-18}$ \quad \\[2mm]
		$n=1$ & $2.9921\cdot 10^{-1}$ & $-1.2100\cdot 10^{-1}$ & $5.4536\cdot
		10^{-3} $ & $2\cdot 10^{-15}$ \\[2mm]
		$n=2$ & $2.5557\cdot 10^{-1}$ & $-1.3611\cdot 10^{-1}$ & $2.1097\cdot
		10^{-2} $ & $5\cdot 10^{-13}$ \\ \hline
	\end{tabular}%
	\caption{The ac case. The values of $\Delta _{n}$, $n=0,1,2$, calculated as per Eq. (%
		\protect\ref{omega2}), for $P(x)=\cos \left( \Omega x\right) $, $\Omega
		=0,2,8,18$. They determine the perturbative shift of chemical potential of
		weakly nonlinear LMs, pursuant to Eq. (\protect\ref{AsEx}). }
	\label{T2}
	\end{center}
\end{table}

To find the asymptotic form of $\Delta _{n}$ for $\Omega \rightarrow \infty $%
, we note that, for an arbitrary polynomial $%
Q_{2m}(x)=a_{2m}x^{2m}+a_{2m-1}x^{2m-1}+\ldots +a_{0}$ of degree $2m$, the
following asymptotic relation holds:
\begin{equation}
\int_{-\infty }^{+\infty }Q_{2m}(x)e^{-2x^{2}}\cos \left( \Omega x\right)
~dx\approx (-1)^{m}\frac{a_{2m}\sqrt{2\pi }\Omega ^{2m}}{2^{4m+1}}e^{-\Omega
	^{2}/8},\quad \Omega \rightarrow \infty.  \label{AsGauss}
\end{equation}%
Since the coefficient in front of the highest-power term in the Hermite
polynomial $H_{n}(x)$ is $2^{n}$, Eq. (\ref{AsGauss}) with $Q_{2m}(x)$
substituted by $H_{n}(x)$ yields
\begin{equation}
\Delta _{n}\approx \frac{\sqrt{2}\Omega ^{4n}}{2^{6n+1}\cdot (n!)^{2}\sqrt{%
		\pi }}e^{-\Omega ^{2}/8},\quad \Omega \rightarrow \infty.
\end{equation}%
Therefore, for sufficiently large $\Omega $, $\Delta _{n}$ {are all positive}
and decay exponentially, which agrees with the data in Table~\ref{T2}.

\begin{figure}[tbp]
	\begin{center}
	\includegraphics[width=0.9\textwidth]{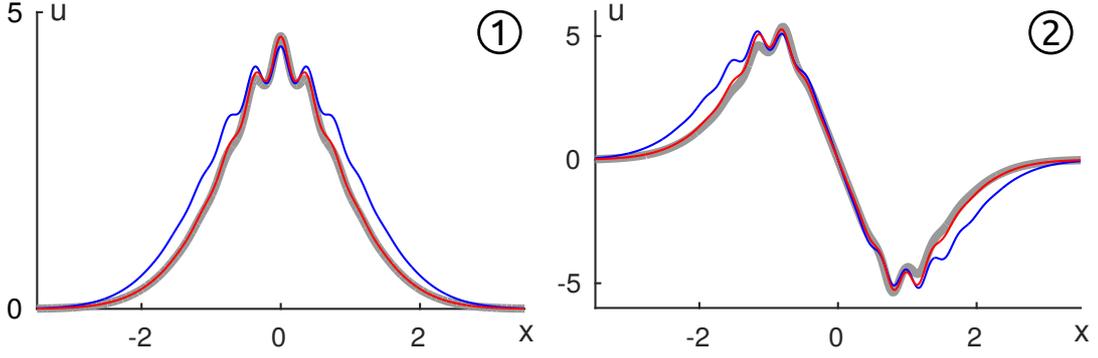}
	\caption{The ac case. Profiles of LMs, for $P(x)=\cos \left( 16x\right) $. (1) Branch $%
		\Gamma _{0}$, $\protect\mu =0$ ($\Delta \protect\mu =1$). (2) Branch $\Gamma
		_{1}$, $\protect\mu =1$ ($\Delta \protect\mu =2$). Gray line: the numerical
		solution, red line: approximation (\protect\ref{AsRel_u}), with $U(x)$
		obtained by numerical solution of Eq. (\protect\ref{U}). Blue line:
		approximation (\protect\ref{u_appr_main}). }
	\label{Fig:Prof}
	\end{center}
\end{figure}

As has been shown above [see Eq.~(\ref{Expand_NZM})], in the model with 
rapidly oscillating pseudopotential and nonzero dc term, the shapes of LMs may
be approximated by the corresponding solutions of the nonlinear HO equation.
However, the asymptotic behavior of LMs in the limit of $\Omega \rightarrow
\infty $ becomes essentially different for the pseudopotential without the
dc term. Indeed, nonlinear mode $u(x)$ belonging to the branch $\Gamma _{n}$
which satisfies Eq.~(\ref{eq1_2}), may be looked for as
\begin{equation}
u(x)=W(x)+R(x)\cos \left( \Omega x\right) +\mathrm{higher}~\mathrm{order}~%
\mathrm{terms},  \label{Asym_hyp}
\end{equation}%
where $W(x)$ and $R(x)$ are slowly varying functions in comparison with $%
\cos \left( \Omega x\right) $. Then one arrives at the balance relation between
the dominating terms:
\begin{equation}
R(x)\Omega ^{2}=\sigma _{1}W^{3}(x).
\end{equation}%
In the next order of $1/\Omega $, one arrives at the equation for $W(x)$,
\begin{equation}
\frac{d^{2}W}{dx^{2}}+(\mu -x^{2})W+\frac{3}{2\Omega ^{2}}W^{5}=0.
\end{equation}%
Introducing a rescaled function, $U(x)=W(x)/\sqrt{\Omega }$, one obtains the
following approximation:
\begin{equation}
u(x)=\sqrt{\Omega }U(x)+\frac{\sigma _{1}}{\sqrt{\Omega }}U^{3}(x)\cos
\Omega x+o\left( \frac{1}{\sqrt{\Omega }}\right) ,  \label{AsRel_u}
\end{equation}%
where $U(x)$ is a localized solution of the equation
\begin{equation}
\frac{d^{2}U}{dx^{2}}+(\mu -x^{2})U+\frac{3}{2}U^{5}=0.  \label{U}
\end{equation}%
While Eq.~(\ref{U}) does not admit any simple analytical solution, its
localized modes can be computed numerically and used in Eq. (\ref{AsRel_u})
to approximate nonlinear modes in the rapidly oscillating pseudopotential
(compare grey and red lines in Fig.~\ref{Fig:Prof}). To obtain the asymptotic
expression for the shapes of nonlinear modes in explicit analytical form,
we define $\Delta \mu =\tilde{\mu}_{n}-\mu $ and rewrite Eq.~(\ref{U}) as
\begin{equation}
\frac{d^{2}U}{dx^{2}}+(\tilde{\mu}_{n}-x^{2})U=-\frac{3}{2}U^{5}+\Delta \mu
U.  \label{U0}
\end{equation}%
If $\Delta \mu $ is not large, one may assume that $U(x)\approx
U_{0}H_{n}(x)e^{-x^{2}/2}$, where $U_{0}$ is some constant which can be
found from the requirement for right-hand side of (\ref{U0}) to be
orthogonal to the kernel of the operator in the left-hand side:
\begin{equation}
\int_{-\infty }^{+\infty }\left( -\frac{3}{2}U^{5}(x)+\Delta \mu U(x)\right)
H_{n}(x)e^{-x^{2}/2}~dx=0,  \label{Orthog}
\end{equation}%
which leads to
\begin{equation}
U_{0}=\left( \frac{2{\mathcal{H}}_{n}^{(2)}\Delta \mu }{3{\mathcal{H}}%
	_{n}^{(6)}}\right) ^{\frac{1}{4}},  \label{Q_0_formula}
\end{equation}%
\begin{equation}
{\mathcal{H}}_{n}^{(2)}\equiv \int_{-\infty }^{+\infty
}H_{n}^{2}(x)e^{-x^{2}}~dx,\quad {\mathcal{H}}_{n}^{(6)}\equiv \int_{-\infty
}^{+\infty }H_{n}^{6}(x)e^{-3x^{2}}~dx.
\end{equation}

Combining the obtained results with Eq. (\ref{AsRel_u}), one finally gets
\begin{equation}
u(x)\approx \left( \frac{2{\mathcal{H}}_{n}^{(2)}\Delta \mu }{3{\mathcal{H}}%
	_{n}^{(6)}}\right) ^{1/4}\sqrt{\Omega }H_{n}(x)e^{-x^{2}/2}+\left( \frac{2{%
		\mathcal{H}}_{n}^{(2)}\Delta \mu }{3{\mathcal{H}}_{n}^{(6)}}\right) ^{3/4}%
\frac{\sigma _{1}}{\sqrt{\Omega }}H_{n}^{3}(x)e^{-3x^{2}/2}\cos \left(
\Omega x\right).  \label{u_appr_main}
\end{equation}%
For the lowest branch $\Gamma _{0}$ Eq. (\ref{u_appr_main}) yields
\begin{equation}
u(x)\approx \left( \frac{2\Delta \mu }{\sqrt{3}}\right) ^{1/4}\sqrt{\Omega }%
e^{-x^{2}/2}+\left( \frac{2\Delta \mu }{\sqrt{3}}\right) ^{3/4}\frac{\sigma
	_{1}}{\sqrt{\Omega }}e^{-3x^{2}/2}\cos \left( \Omega x\right).
\end{equation}%
For the next branch, $\Gamma _{1}$, one obtains
\begin{equation}
u(x)\approx \left( \frac{216\Delta \mu }{15\sqrt{3}}\right) ^{1/4}\sqrt{%
	\Omega }xe^{-x^{2}/2}+\left( \frac{216\Delta \mu }{15\sqrt{3}}\right) ^{3/4}%
\frac{\sigma _{1}}{\sqrt{\Omega }}x^{3}e^{-3x^{2}/2}\cos \left( \Omega
x\right).
\end{equation}%
The comparison between the numerical solution and approximations (\ref%
{AsRel_u}) and (\ref{u_appr_main}) is illustrated by the Fig.~\ref{Fig:Prof}, for $\Omega =16$. The figure shows that approximation (\ref{AsRel_u})
describes the numerical solution very well. Approximation (\ref{u_appr_main}%
) is good too, even if $\Delta \mu $ is not small (compare grey and blue
lines in Fig.~\ref{Fig:Prof}, where $\Delta \mu =1$ and $2$ in the left and
right panels, respectively).

\subsubsection{Stability of localized modes}

\label{Gamma_stab}

In the small-amplitude limit, the stability of LMs is determined by the
eigenvalues of matrix $\tilde{M}_{n}$, see Sect.~\ref{SmAmpl}. Using Maple,
one can compute these eigenvalues with any necessary accuracy. The results
for two modulation frequencies ($\Omega =8$ and $16$) are summarized in
Tables~\ref{T4}-\ref{T5}. In these tables, $n$ is the index of the branch $%
\Gamma _{n}$ (for instance, $n=2$ means that the branch $\Gamma _{2}$, which
starts from $\tilde{\mu}_{2}=2\cdot 2+1=5$, is under consideration), while $%
k $, running from $0$ to $n-1$, enumerates double eigenvalues $\tilde{\Lambda%
}_{k}=4(n-k)^{2}$. Each cell with $n\geq k$ contains either letter
``C" or two real numbers. These numbers
are the real eigenvalues $\gamma _{1,2}$ of the matrix $\tilde{M}_{n}$,
whereas ``C" means that both eigenvalues
of $\tilde{M}_{n}$ are complex.
For $n<k$ the double eigenvalues do not exist (dashes at the corresponding
positions in the table).

\begin{table}[tbp]
	\begin{center}
	\begin{tabular}{ccccccccccc}
		& \phantom{111} & $n=1$ &  & $n=2$ &  & $n=3$ &  & $n=4$ &  & $n=5$ \\ \hline
		$k=0$ &  & $%
		\begin{array}{c}
		0.132 \\
		-0.028%
		\end{array}%
		$ &  & $%
		\begin{array}{c}
		-0.012 \\
		0.203%
		\end{array}%
		$ &  & C &  & $%
		\begin{array}{c}
		0.424 \\
		-0.828%
		\end{array}%
		$ &  & $%
		\begin{array}{c}
		-0.845 \\
		-0.660%
		\end{array}%
		$ \\ \hline
		$k=1$ &  & -- &  & $%
		\begin{array}{c}
		-0.176 \\
		0.159%
		\end{array}%
		$ &  & $%
		\begin{array}{c}
		0.432 \\
		-0.190%
		\end{array}%
		$ &  & $%
		\begin{array}{c}
		-0.486 \\
		0.048%
		\end{array}%
		$ &  & $%
		\begin{array}{c}
		0.356 \\
		0.944%
		\end{array}%
		$ \\ \hline
		$k=2$ &  & -- &  & -- &  & $%
		\begin{array}{c}
		-0.068 \\
		-0.132%
		\end{array}%
		$ &  & $%
		\begin{array}{c}
		-0.206 \\
		0.196%
		\end{array}%
		$ &  & $%
		\begin{array}{c}
		0.531 \\
		-0.600%
		\end{array}%
		$ \\ \hline
		$k=3$ &  & -- &  & -- &  & -- &  & C &  & $%
		\begin{array}{c}
		0.166 \\
		-0.122%
		\end{array}%
		$ \\ \hline
		$k=4$ &  & -- &  & -- &  & -- &  & -- &  & C \\ \hline
	\end{tabular}%
	\caption{Eigenvalues of matrix $\tilde{M}$,\ in the case of $P(x)=\cos
		\left( 8x\right) $. Here $n$ is the index of branch $\Gamma _{n}$, and $k$
		enumerates double eigenvalues $\tilde{\Lambda}_{k}=4(n-k)^{2}$. Each cell
		with $n>k$ contains either letter ``C" or
		two real numbers. These numbers are real eigenvalues $\protect\gamma _{1,2}$
		of matrix $\tilde{M}_{n}$, whereas letter ``C" means that eigenvalues
		of $\tilde{M}_{n}$ are complex.}
	\label{T4}
	\end{center}
\end{table}

\begin{table}[tbp]
\begin{center}
	\begin{tabular}{ccccccccccc}
		& \phantom{111} & $n=1$ &  & $n=2$ &  & $n=3$ &  & $n=4$ &  & $n=5$ \\ \hline
		$k=0$ &  & $%
		\begin{array}{c}
		-2.0\cdot 10^{-11} \\
		5.7\cdot 10^{-10}%
		\end{array}%
		$ &  & $%
		\begin{array}{c}
		-7.6\cdot 10^{-9} \\
		-9.2\cdot 10^{-7}%
		\end{array}%
		$ &  & $%
		\begin{array}{c}
		-8.7\cdot 10^{-7} \\
		2.2\cdot 10^{-4}%
		\end{array}%
		$ &  & $%
		\begin{array}{c}
		-4.0\cdot 10^{-5} \\
		-1.2\cdot 10^{-2}%
		\end{array}%
		$ &  & $%
		\begin{array}{c}
		-9.5\cdot 10^{-4} \\
		1.5\cdot 10^{-1}%
		\end{array}%
		$ \\ \hline
		$k=1$ &  & -- &  & $%
		\begin{array}{c}
		-4.2\cdot 10^{-9} \\
		7.3\cdot 10^{-8}%
		\end{array}%
		$ &  & $%
		\begin{array}{c}
		-5.8\cdot 10^{-7} \\
		-3.6\cdot 10^{-5}%
		\end{array}%
		$ &  & $%
		\begin{array}{c}
		-3.0\cdot 10^{-5} \\
		3.3\cdot 10^{-3}%
		\end{array}%
		$ &  & $%
		\begin{array}{c}
		-7.6\cdot 10^{-4} \\
		-6.8\cdot 10^{-2}%
		\end{array}%
		$ \\ \hline
		$k=2$ &  & -- &  & -- &  & $%
		\begin{array}{c}
		-3.4\cdot 10^{-7} \\
		3.9\cdot 10^{-6}%
		\end{array}%
		$ &  & $%
		\begin{array}{c}
		-2.0\cdot 10^{-5} \\
		-6.8\cdot 10^{-4}%
		\end{array}%
		$ &  & $%
		\begin{array}{c}
		-5.7\cdot 10^{-4} \\
		2.6\cdot 10^{-2}%
		\end{array}%
		$ \\ \hline
		$k=3$ &  & -- &  & -- &  & -- &  & $%
		\begin{array}{c}
		-1.3\cdot 10^{-5} \\
		1.0\cdot 10^{-4}%
		\end{array}%
		$ &  & $%
		\begin{array}{c}
		-3.6\cdot 10^{-4} \\
		-6.8\cdot 10^{-3}%
		\end{array}%
		$ \\ \hline
		$k=4$ &  & -- &  & -- &  & -- &  & -- &  & $%
		\begin{array}{c}
		-2.7\cdot 10^{-4} \\
		1.5\cdot 10^{-3}%
		\end{array}%
		$ \\ \hline
	\end{tabular}%
	\caption{Eigenvalues of matrix $\tilde{M}_{n}$ for $P(x)=\cos \left(
		16x\right) $. The notation is the same as in Table \protect\ref{T4}.}
	\label{T5}
	\end{center}
\end{table}

To be specific, let us describe in detail the branch $\Gamma _{2}$ in the case
of $P(x)=\cos \left( 8x\right) $ (i.e., $n=2$ in
Table~\ref{T4}). In this case, there are two double eigenvalues in the
spectrum, $\tilde{\Lambda}_{0}=16$ and $\tilde{\Lambda}_{1}=4$. According to
Table~\ref{T4} and Eq. (\ref{Lambda_12_as}), they split as
\begin{equation}
\Lambda _{0}^{(1)}=16-0.012\cdot \varepsilon ^{2}+\ldots ,\quad \Lambda
_{0}^{(2)}=16+0.203\cdot \varepsilon ^{2}\ldots ~
\end{equation}%
and
\begin{equation}
\Lambda _{1}^{(1)}=4-0.176\cdot \varepsilon ^{2}+\ldots ,\quad \Lambda
_{1}^{(2)}=4+0.159\cdot \varepsilon ^{2}\ldots ~,
\end{equation}%
hence the small-amplitude nonlinear LMs belonging to the branch $\Gamma _{2}$
are stable in this case. The situation is different for $\Gamma _{3,4,5}$,
since for each of these branches the bifurcation of a complex-conjugate pair
occurs: for $n=3$ the eigenvalue $\tilde{\Lambda}_{0}=36$ splits into
complex eigenvalues,while for $n=4$ and $n=5$ this takes place for $\tilde{%
	\Lambda}_{2}=4$. Therefore, for $\Omega =8$ the small-amplitude LMs are
stable in the branches $\Gamma _{0}$, $\Gamma _{1}$ and $\Gamma _{2}$, but
unstable in $\Gamma _{3}$, $\Gamma _{4}$ and $\Gamma _{5}$. Table~\ref{T5}
is produced for $\Omega =18$, implying that the small-amplitude LMs are
stable for \textit{all} the branches, $\Gamma _{0,\ldots,5}$.

To explain the different stability of small-amplitude LMs with different
spatial frequencies $\Omega $, we consider the behavior of the eigenvalues of $%
\tilde{M}_{n}$ at $\Omega \rightarrow \infty $. Using explicit results given
by Eqs. (\ref{11})--(\ref{22}) and asymptotic relation (\ref{AsGauss}), we
obtain
\begin{align}
\langle M_{n}\tilde{u}_{k},\tilde{u}_{k}\rangle & \sim -\frac{2\sqrt{2}%
	(n-k)\Omega ^{4n}}{2^{6n}\cdot (n!)^{2}\sqrt{\pi }}e^{-\Omega ^{2}/8}, \\%
[2mm]
\langle M_{n}\tilde{u}_{k},\tilde{u}_{2n-k}\rangle & =-\langle M_{n}\tilde{u}%
_{2n-k},\tilde{u}_{k}\rangle \sim \frac{2\sqrt{2}(n-k)\Omega ^{4n}}{2^{6n}n!%
	\sqrt{k!(2n-k)!}\sqrt{\pi }}e^{-\Omega ^{2}/8}, \\
\langle M_{n}\tilde{u}_{2n-k},\tilde{u}_{2n-k}\rangle & \sim -\frac{4\sqrt{2}%
	(n-k)\Omega ^{2(3n-k)}}{\sqrt{\pi }2^{3(3n-k)}n!(2n-k)!}e^{-\Omega ^{2}/8}.
\end{align}%
These relations imply that
\begin{equation}
\tilde{M}_{n}=\left(
\begin{array}{cc}
O(\Omega ^{4n}e^{-\Omega ^{2}/8}) & O(\Omega ^{4n}e^{-\Omega ^{2}/8}) \\%
[2mm]
O(\Omega ^{4n}e^{-\Omega ^{2}/8}) & O(\Omega ^{2(3n-k)}e^{-\Omega ^{2}/8})%
\end{array}%
\right) ,\quad \Omega \rightarrow \infty
\end{equation}%
All elements of the matrix $\tilde{M}_{n}$ are of the same order, except for the
one in the right lower corner which is of greater order due to $n>k$. This
means that, for\textit{\ }$\Omega $\textit{\ }large enough, eigenvalues of $%
\tilde{M}_{n}$\textit{\ }are real, hence the nonlinear LMs from branch $%
\Gamma _{n}$ for arbitrary $n$ are stable, at least in the small-amplitude
limit. This explains the difference between the results displayed in Tables~%
\ref{T4} and \ref{T5}: the increase of the spatial frequency from $\Omega =8$
to $\Omega =16$ results in stabilization of some small-amplitude LMs which
were originally unstable.

\begin{figure}[tbp]
	\begin{center}
	\includegraphics[width=0.8\textwidth]{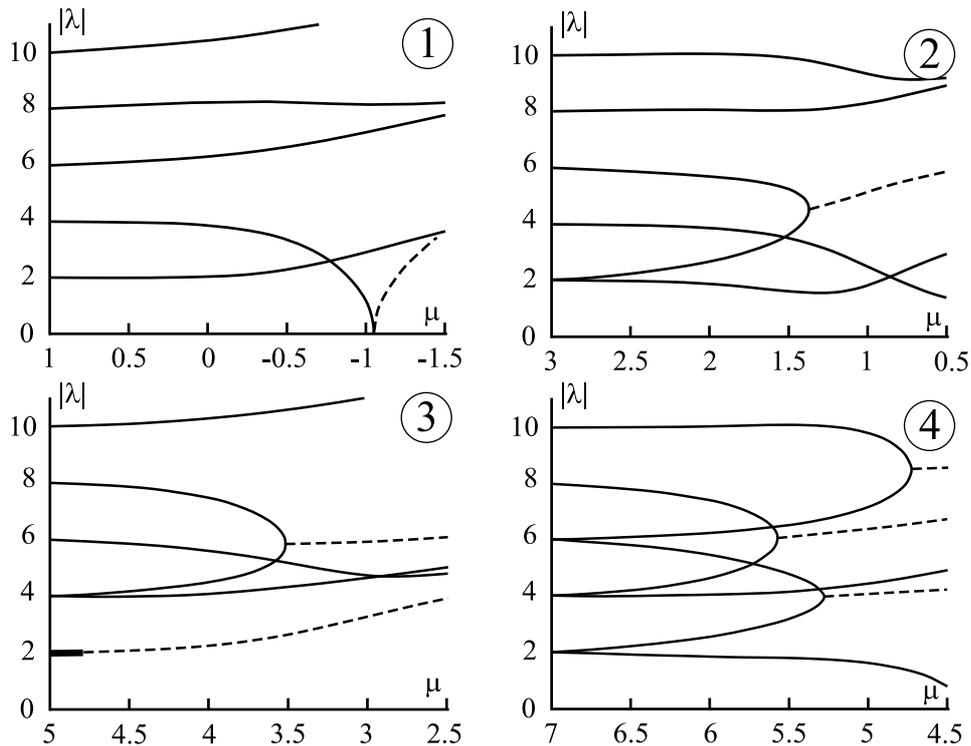}
	\caption{The evolution of eigenvalues of (\protect\ref{EigProb}) as $\protect%
		\mu $ varies, for $P(x)=\cos \left( 16x\right) $. Only eigenvalues with $%
		\mathrm{Im}~\protect\lambda \geq 0$ are shown. Solid line: pure imaginary
		eigenvalues corresponding to stable LMs; dashed line: eigenvalues with a
		nonzero real part. (1) Branch $\Gamma _{0}$. No double eigenvalues exist at $%
		\protect\mu =1$, all the eigenvalues being pure imaginary and simple, until
		the collision at $\mu\approx -1.049$. After the collision a
		pair of real eigenvalues emerges. (2) Branch $\Gamma _{1}$. At $\protect\mu %
		=3$ there exists one double eigenvalue, $\protect\lambda =2i$. According to
		the prediction of the asymptotic analysis, it splits into a pair of pure
		imaginary simple eigenvalues. The collision of eigenvalues occurs at
		$\mu\approx 1.372$, and a pair of eigenvalues with nonzero
		real parts emerges. (3) Branch $\Gamma _{2}$. At $\protect\mu =5$, there
		exists two pairs of double eigenvalues, $\protect\lambda =2i$ and $\protect%
		\lambda =4i$. Both of them split into pairs of pure imaginary eigenvalues.
		The former pair collides again at $\mu\approx 4.801$ and
		transforms into a pair of complex eigenvalues with nonzero real parts. At
		$\mu\approx 3.519$ , another collision of eigenvalues
		occurs. (4) Branch $\Gamma _{3}$. At $\protect\mu =7$, there exist three
		pairs of double eigenvalues. The first collision of pure imaginary
		eigenvalues takes place at $\mu\approx 5.571$, then two more
		collisions occur. They all result in eigenvalues with nonzero real parts.}
	\label{CollEig}
	\end{center}
\end{figure}

Rigorously speaking, the results for the linear stability of the
small-amplitude LMs are asymptotic, and, while moving along a branch $\Gamma
_{n}$ to the region of moderate or large amplitudes, the nonlinear LMs may
change the stability.
To illustrate the persistence of the asymptotic predictions for the
(in)stability of the nonlinear LMs, in Fig.~\ref{CollEig} we plot
numerically obtained dependencies of eigenvalues $\lambda $ on $\mu $ for
branches $\Gamma _{0,1,2,3}$ in the case of $P(x)=\cos \left( 16x\right) $.
Each shown branch is stable near the point of its emergence, losing
stability at some threshold value of $\mu $ (which is different for
different branches).

\begin{figure}[tbp]
	\begin{center}
	\includegraphics[width=0.8\textwidth]{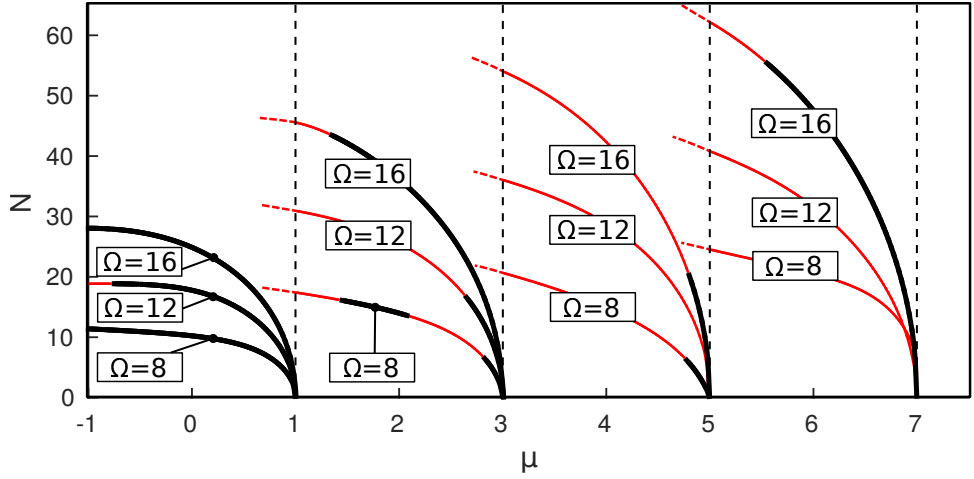}%
	\caption{The ac case. $N(\protect\mu )$ curves for $P(x)=\cos \left( \Omega x\right) $, $%
		\Omega =8,12,16$. Branches $\Gamma _{0,1,2,3}$ are presented. Black bold
		(red thin) lines correspond to stable (unstable) LMs.}
	\label{Fig_ZM_branches}
	\end{center}
\end{figure}

Numerically generated $N(\mu )$ curves for $\Omega =8$, $\Omega =12$, and $%
\Omega =16$ are plotted in Fig.~\ref{Fig_ZM_branches}. It follows from these
plots that the branch $\Gamma _{0}$ is the ``most
stable'' one for all the three values of $\Omega $. One
also notices that, for the greatest value of $\Omega $ (i.e., $\Omega =16$)
there exists a ``stability window'' in a
vicinity of the bifurcations for all branches $\Gamma _{0,1,2,3}$, in
agreement with the asymptotic results presented above.

\begin{figure}[tbp]
	\begin{center}
	\includegraphics[width=0.8\textwidth]{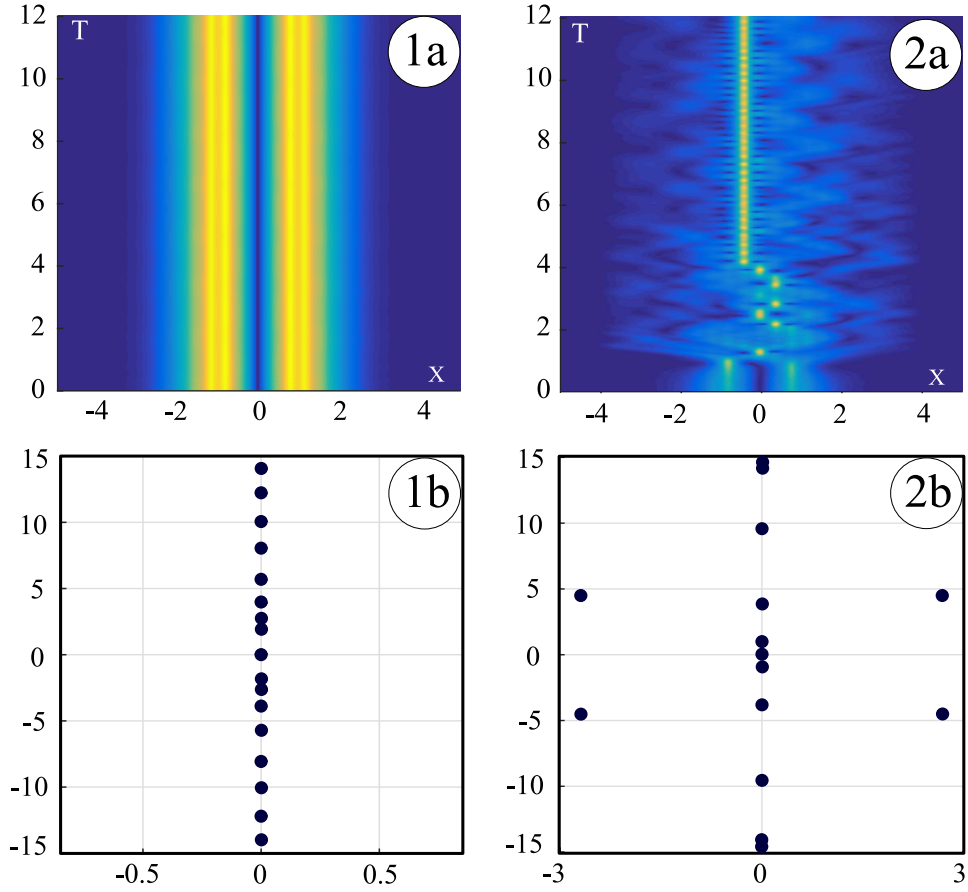}
	\caption{The ac case, $P(x)=\cos \left( 16x\right) $. (1a): The evolution of
		a stable nonlinear mode belonging to branch $\Gamma _{1}$, with $\protect\mu %
		=2$. (1b): the spectrum of eigenvalues $\protect\lambda $ produced by
		numerical solution of Eq. (\protect\ref{EigProb}) associated with this mode.
		(2a): The evolution of an unstable mode belonging to branch $\Gamma _{1}$,
		with $\protect\mu =0$; (2b) the corresponding spectrum. }
	\label{Fig_Comp_ZM}
	\end{center}
\end{figure}

Numerical study of the temporal evolution of LMs in the framework of
time-dependent GPE~(\ref{1D_GP_12}) confirms the predictions of the
linear-stability analysis, see \textcolor{black}{Fig.~\ref{Fig_Comp_ZM}}.
As an example of a distinctive pattern of dynamical behavior of unstable
modes, in \textcolor{black}{Fig.~\ref{Fig_Comp_ZM}~(2a)} we display the transformation of an unstable solution into a
pulsating object localized over one period of the lattice pseudopotential.

\section{Conclusion}

\label{Concl}

In the paper, we have systematically studied nonlinear localized modes (LMs)
of the one-dimensional Gross-Pitaevskii/nonlinear Schr\"{o}dinger equation
with the  harmonic-oscillator (HO) trapping potential $V(x)$ and nonlinear
lattice pseudopotential $P(x)$, which is an periodic function oscillating
with spatial frequency $\Omega $. This equation  describes a cigar-shaped cloud of BEC confined by the magnetic
trap and manipulated by the optically-induced Feshbach resonance with the
periodically modulated local strength. The model is characterized by the
interplay between two spatial scales: the characteristic size of the HO trap
and the period of the nonlinear lattice pseudopotential. This factor
essentially enriches the variety of nonlinear modes in this model, in
comparison with the well-studied case of the spatially uniform nonlinearity.
We have obtained analytical results of the shape of nonlinear LMs and on
their stability, in the limits of small-amplitude solutions and rapidly
oscillating pseudopotential. The validity and persistence of the asymptotic
predictions has been corroborated by systematically presented numerical
findings.

To conclude the paper, it is relevant, once again, \textcolor{black}{to} highlight its   main
results. It was found that there exist two types of branches of LMs, \textit{viz}.,
ones with and without linear counterpart. This is especially interesting in
view of the fact that no modes without linear counterpart exists in the
model with the uniform nonlinearity. However, all these ``exotic''
nonlinear modes, disconnected from the linear limit, are unstable.
As concerns nonlinear LMs bifurcating from eigenstates of the underlying
linear problem, their properties are essentially different depending on the
presence of the dc term (nonzero mean value) in the periodic
pseudopotential. \textcolor{black}{If} the mean is nonzero, then properties of nonlinear modes
in a rapidly oscillating pseudopotential may be approximated using solutions
for the standard HO model with constant nonlinearity. However, the reduction
to the standard model with constant nonlinearity does not work for the
system in a rapidly oscillating pseudopotential without the dc term. Most
interestingly, in this case we have found that the rapidly oscillating
pseudopotential can stabilize small-amplitude LMs belonging to higher
families (excited states, which are unstable in the model including the dc
term). Specifically, for any given branch with index $n$, there exists a
threshold value of the spatial frequency, $\Omega _{n}$, such that the
small-amplitude solutions belonging to this branch are stable for $\Omega
>\Omega _{n}$.

As a continuation of this work, it may be interesting to extend the analysis
to the case of the interplay between  the spatially periodic pseudopotential and  the expulsive (anti-HO) potential. A more challenging possibility is to
consider a two-dimensional model featuring the combination of an isotropic
HO trapping potential and two-dimensional lattice quasipotential.

\section*{Acknowledgments}
The research of G.L.A., M.E.L. and D.A.Z. was supported by Russian Science
Foundation (Grant No. 17-11-01004). The work of B.A.M. is supported, in
part, by the joint program in physics between NSF and Binational (US-Israel)
Science Foundation through project No. 2015616, and by the Israel Science
Foundation through Grant No. 1286/17. The research of D.A.Z.  is also supported by Government of Russian Federation (Grant 08-08).

\end{document}